\documentclass[onecolumn,showpacs,preprintnumbers,amsmath,amssymb]{revtex4}
\usepackage{dcolumn}% Align table columns on decimal point
\usepackage{bm}% bold math
\usepackage{graphicx}%
\usepackage{epsfig}
\begin{document}
\title{Truncated generalized coherent states}
\author{Filippo Giraldi }
\email{giraldi@ukzn.ac.za} \affiliation{School of Chemistry and Physics, 
University of KwaZulu-Natal, Westville Campus, Durban 4000, South Africa.}
\author{Francesco Mainardi}
\email{francesco.mainardi@bo.infn.it}
\affiliation{University of Bologna and INFN
Department of Physics \& Astronomy, Italy.}


\begin{abstract}
A generalization of the canonical coherent states of a quantum harmonic oscillator has been performed by requiring the conditions of normalizability, continuity in the label and resolution of the identity operator with a positive weight function. Relying on this approach, in the present scenario coherent states are generalized over the canonical or finite dimensional Fock space of the harmonic oscillator. A class of generalized coherent states is determined such that the corresponding distributions of the number of excitations depart from the Poisson statistics according to combinations of stretched exponential decays, power laws and logarithmic forms. The analysis of the Mandel parameter shows that the generalized coherent states exhibit (non-classical) sub-Poissonian or super-Poissonian statistics of the number of excitations, based on the realization of determined constraints. Mittag-Leffler and Wright generalized coherent states are analyzed as particular cases. 
\end{abstract}

\maketitle

\section{Introduction}\label{1}

Coherent states (CSs) are fundamental quantum states of an harmonic oscillator which find the most various applications in mathematical physics and quantum optics, to name but a few. See Refs. \cite{Sch1926,Glauber1963,Klauder1963,Klauder1985,Perelomov1986,Klauder1995,Loudon1983,WallsMilbourn1994,MandelWolf1995,SCSsSanders2012}, to name but a few. CSs were originally studied by Schr\"odinger analyzing the time evolution of Gaussian wave packets \cite{Sch1926} for a quantum harmonic oscillator. The mean value of these wave packets evolves as a classical harmonic oscillator while the spread is fixed. Due to these properties, CSs are considered to be quasi-classical states. Glauber determined CSs by requiring the condition of minimum uncertainty \cite{Glauber1963}. In this way, CSs are introduced as the eigenstates of the annihilation operator of the quantum harmonic oscillator.

During the last decades, various generalizations of CSs are proposed in literature. See Refs. \cite{Glauber1963,TitulaerGlauber1966,Bialynicki1968,Spiridonov1995,BarutGirardelloCSs1971}, to name but a few. Generalization of CSs is performed by requiring the condition of full coherence to be fulfilled \cite{Glauber1963,TitulaerGlauber1966,Bialynicki1968,Spiridonov1995}. This approach provides quantum states which, in general, are not minimum-uncertainty wave packets \cite{Stolter1971}. An algebra generalization of CSs allows to interpret CSs as orbits of minimum uncertainty states \cite{BarutGirardelloCSs1971}. A generalization of CSs is provided by Klauder \cite{Klauder1963} requiring the following conditions to hold: normalizability; continuity in the label which characterized these states; resolution of the identity operator with a positive weight function (completeness condition). This approach allows the definition of special generalized coherent states (GCSs) which are referred to as Mittag-Leffler and Wright GCSs. These special states are defined by generalizing the Gamma function, which appears in the representation of the CSs in the Fock basis of the quantum harmonic oscillator (specifically, the factorial term), with the Mittag-Leffler and Wright functions \cite{MLcoher,MLcoherJMP,weightedPoissBalak,Wcoher}. This route reproduces the generalization of the Poisson distribution with weighted Poisson distributions. Another way of generalizing CSs is based on fractional Poisson distributions \cite{MLLaskinJMP2009,GieselVetter2021}. The corresponding GCSs are obtained by considering the representation of the CSs in the Fock basis and introducing fractional powers of the label. These GCSs satisfy the conditions of normalizability, continuity in the label, and resolution of the identity operator with a positive weight function and, additionally, are eigenstates of the annihilation operator \cite{GieselVetter2021}.

Truncated CSs are introduced by considering the Fock space of the quantum harmonic oscillator to be finite-dimensional. This approach was originally introduced to develop an appropriate definition of the phase operator. Refer to \cite{finiteDimH1,finiteDimH2,finiteDimH3,TCSsdefs1990,Truncosc1992,finiteDimCoher1,Truncosc21994,Truncosc1997,Truncosc2011,TruncoscXiv2014,Truncosc2020} for details, to name but a few. In Ref. \cite{Truncosc1992}, truncated CSs are defined by evaluating the displacement operator in the finite-dimensional Fock space and applying this operator to the vacuum state of the harmonic oscillator. In Refs. \cite{finiteDimCoher1,Truncosc2020}, truncated CSs are defined by truncating the series which represents the CSs in the infinite dimensional Fock space. In this way, truncated CSs are defined via the truncated exponential function \cite{finiteDimCoher1,Truncosc2020}. Truncated CSs exhibit novel peculiarities with respect to the ones which characterize canonical CSs. In fact, truncated CSs show further squeezing properties and bunching or anti-bunching effects, to name but few \cite{finiteDimCoher1,Truncosc2020}.

 As a continuation of the above-described scenario, here, we rely on the analysis which is performed in Refs. \cite{finiteDimCoher1,MLcoher,Wcoher,Truncosc2020}, and introduce the truncated generalized coherent states of the truncated harmonic oscillator by considering the Klauder's generalization of CSs over the finite dimensional Fock space. Particularly, we study the distribution of the number of excitations and the non-classical properties of truncated Mittag-Leffler and Wright GCSs or further generalizations. The paper is organized as follows. Section \ref{2} is devoted to the description of the CSs, Mittag-Leffler and Wright GCSs and further generalizations of CSs. Truncated  Mittag-Leffler, Wright and GCSs are introduced in Section \ref{3}. The Mandel parameter of GCSs or truncated GCSs, is evaluated in Section \ref{4} for small or large values of the label. Summary of the results and conclusions are provided in Section \ref{5}. Details of the calculations are given in the Appendix.

\section{Generalized coherent states }\label{2}

Canonical CSs are defined in the Fock space of the quantum harmonic oscillator as follows:
\begin{eqnarray}
|z\rangle= \exp \left(-\frac{\left|z\right|^2}{2}\right)\sum_{n=0}^{\infty}
\frac{z^n}{\sqrt{n!}}|n\rangle. \label{expCoherent1}
\end{eqnarray}
for every nonvanishing complex value of the variable $z$, i.e., for every $z\in \mathbb{C}\setminus \left\{0\right\}$. For vanishing value of the label, $z=0$, the CS coincides with the vacuum state $|0\rangle$. The state kets $|0\rangle$, $|1\rangle$, \ldots, are the eigenstates of the number operator $\hat{N}$ with the eigenvalue $0,1,\ldots$, respectively, i.e, $\hat{N} |n\rangle =n |n\rangle $ for every natural value of the variable $n$, i.e., $n \in \mathbb{N}$. These states form an orthonormal and complete set which constitutes the Fock basis of the Hilbert space $\mathcal{H}$ of the quantum harmonic oscillator. Every of these states is related to the vacuum state $|0\rangle$ via the creation operator $a^{\dagger}$ in the following way: $|n\rangle =\left(a^{\dagger}\right)^n |0\rangle/\sqrt{n!}$, for every $n \in \mathbb{N}$. The number operator is defined in terms of the creation operator $a^{\dagger}$ and annihilation operator $a$ as follows: $\hat{N}=a^{\dagger} a$, and the Hamiltonian operator $H$ of the harmonic oscillator is $H=\hbar \omega \left(\hat{N}+1/2\right)$, where $\hbar$ is the Planck constant and $\omega$ is the frequency of the harmonic oscillator. As mentioned above, the CSs (\ref{expCoherent1}) are eigenstates of the annihilation operator, $a|z\rangle=z|z\rangle$ for every $z\in \mathbb{C}$, and achieve minimum uncertainty condition \cite{Glauber1963}. The set of CSs is overcomplete and provide the following resolution of the identity operator $I$ acting over the Hilbert space $\mathcal{H}$ of the harmonic oscillator:
\begin{eqnarray}
\pi^{-1}\int_{\mathbb{R}^2}|z \rangle\langle z| d^2 z=I,
\label{completnessCC}
\end{eqnarray}
 where $d^2 z=d\operatorname{Re}z d\operatorname{Im}z$. No couple of CSs which are orthogonal to each others exists,
\begin{eqnarray}
\langle z_1||z_2\rangle= \exp \left(
z_1^{\ast}z_2-\frac{|z_1|^2+|z_2|^2}{2}\right). \label{orthogonalityCC}
\end{eqnarray}
However, CSs are almost orthogonal, $\langle z_1||z_2\rangle \simeq 0$, in case $|z_1-z_2|\gg 1$. The probability $p\left(n, z\right)$ that the CS $|z\rangle$ consists in $n$ excitations, i.e., the state $|n\rangle$, is given by the Poissonian distribution, 
\begin{eqnarray}
p\left(n, z\right)= \frac{|z|^{2n}\exp\left(-|z|^2\right)
}{n!}. \label{pCC}
\end{eqnarray}

The three conditions which are required by Klauder \cite{Klauder1963} for the definition of the set of GCSs, $\left\{|\zeta \rangle, \,\,\forall\,\, \zeta \in \mathbb{C}\right\}$, are the following: GCSs are normalizable, $\langle \zeta||\zeta \rangle=1$; GCSs are continuous in the label, $|\zeta-\zeta^{\prime}|\to 0\Rightarrow ||\zeta \rangle-|\zeta^{\prime} \rangle||\to 0$; and GCSs provide a resolution of the identity operator $I$ with a positive weight function,
\begin{eqnarray}
\int_{\mathbb{R}^2}U\left(|\zeta|^2\right)|\zeta \rangle\langle \zeta| d^2 \zeta=I,
\label{completnessC}
\end{eqnarray}
where $d^2 \zeta=d\operatorname{Re}\zeta  d\operatorname{Im}\zeta$. This relation corresponds to the completeness property of the GCSs. 
Particularly, for canonical CSs, the weight function is $U\left(|\zeta|^2\right)=\pi^{-1}$, for every $\zeta \in \mathbb{C}$.

The Mittag-Leffler GCSs are defined from the canonical form (\ref{expCoherent1}) by generalizing the term $n!$ with the Gamma function $\Gamma\left(\alpha n+\beta\right)$ for every $n \in \mathbb{N}$, and for every $\alpha, \beta>0$. In this way, the following generalization of CSs is obtained in terms of the Mittag-Leffler function $E_{\alpha,\beta}\left(z\right)$:
\begin{eqnarray}
|z;\alpha,\beta\rangle= \left[E_{\alpha,\beta}\left(\left|z\right|^2\right)\right]^{-1/2} \sum_{n=0}^{\infty}
\frac{z^n }{\sqrt{\Gamma\left(\alpha n+\beta\right)}}|n\rangle, \hspace{1em}\label{MLCoherent1}
\end{eqnarray}
for every $z\in \mathbb{C}\setminus \left\{0\right\}$ and 
$\alpha,\beta>0$, while $|0;\alpha,\beta\rangle=|0\rangle$ for every $\alpha,\beta>0$. The Mittag-Leffler function $E_{\alpha,\beta}\left(z\right)$ is defined by the expression below \cite{ML1,ML2,ML3}, 
\begin{eqnarray}
E_{\alpha,\beta}\left(z\right)=\sum_{n=0}^{\infty}\frac{z^n}{\Gamma\left(\alpha n+\beta\right)}, \label{MLabdef}
\end{eqnarray}
for every $z, \beta \in \mathbb{C}$, and $\operatorname{Re} \alpha >0$. In the present scenario, we consider uniquely positive values of the involved parameters, $\alpha,\beta>0$. In Ref. \cite{MLcoher}, it is shown that the Mittag-Leffler GCSs fulfill the three above-mentioned conditions. Particularly, the following resolution of the identity operator is provided by the Mittag-Leffler GCSs:
\begin{eqnarray}
\int_{\mathbb{R}^2}U^{\left(ML\right)}_{\alpha,\beta}\left(\left|z\right|^2\right)|z;\alpha,\beta \rangle\langle z;\alpha,\beta| d^2 z=I,
\label{completnessMLCS}
\end{eqnarray}
 where $d^2 z=d\operatorname{Re}z  d\operatorname{Im}z$. The corresponding weight function is given by the expression below, 
\begin{eqnarray}
U^{\left(ML\right)}_{\alpha,\beta}\left(u\right)=
\frac{u^{\left(\beta/\alpha\right)-1}\exp\left(-u^{1/\alpha}\right)E_{\alpha,\beta}\left(u\right)}{\pi \alpha },
\label{WeightMLCS}
\end{eqnarray}
for every $u>0$. Two Mittag-Leffler GCSs $|z_1;\alpha,\beta\rangle$ and $|z_2;\alpha,\beta\rangle$ are orthogonal to each others,
\begin{eqnarray}
&&\langle z_1;\alpha,\beta||z_2;\alpha,\beta\rangle= \left[E_{\alpha,\beta}\left(\left|z_1\right|^2\right)E_{\alpha,\beta}\left(\left|z_2\right|^2\right)\right]^{-1/2}
E_{\alpha,\beta}\left(z_1^{\ast}z_2\right)
=0, \label{orthogonalityMLCSs}
\end{eqnarray}
in case the complex number $\left(z^{\ast}_1z_2\right)$ is a zero of the Mittag-Leffler function, $E_{\alpha,\beta}\left(z_1^{\ast}z_2\right)=0$. The set of these zeros is countable. Therefore, two Mittag-Leffler GCSs are orthogonal uniquely on a set of vanishing measure. Refer to \cite{MLcoher} for details. The probability $p_{\alpha,\beta}\left(n, z\right)$ that the Mittag-Leffler GCS $|z;\alpha,\beta\rangle$ consists in $n$ excitations is given by the expression below,
\begin{eqnarray}
p_{\alpha,\beta}\left(n, z\right)=\frac{\left|z\right|^{2n} }{E_{\alpha,\beta}\left(\left|z\right|^2\right)\Gamma\left(\alpha n + \beta\right) }, \label{pML}
\end{eqnarray}
for every $n \in \mathbb{N}$. This probability exhibits the following asymptotic behavior for large values of the number of excitations:
\begin{eqnarray}
p_{\alpha,\beta}\left(n, z\right)\sim
\frac{\left|z\right|^{2n} e^{\alpha n}
\left(\alpha n\right)^{-\alpha n
-\beta+\left(1/2\right)}}{\left(2 \pi\right)^{1/2} 
E_{\alpha,\beta}\left(\left|z\right|^2\right) }
, \label{pEasympt}
\end{eqnarray}
as $n \gg 1$.

The Wright GCSs are defined from the canonical form (\ref{expCoherent1}) by substituting the term $n!$ with the term $n! \Gamma\left(\lambda n + \mu\right)$, for every $n \in \mathbb{N}$, and $\lambda, \mu>0$. In this way, the Wright GCSs are defined via the Wright function $W_{\lambda,\mu}\left(z\right)$ as follows \cite{Wrightdef}:
\begin{eqnarray}
|z;\lambda,\mu\rangle= \left[W_{\lambda,\mu}\left(\left|z\right|^2\right)\right]^{-1/2} \sum_{n=0}^{\infty}
\frac{z^n }{\sqrt{n! \Gamma\left(\lambda n + \mu\right)}}|n\rangle. \label{WCoherent1}
\end{eqnarray}
for every $z\in \mathbb{C}\setminus \left\{0\right\}$ and 
$\lambda,\mu>0$, while $|0;\lambda,\mu\rangle=|0\rangle$ for every $\lambda,\mu>0$. The Wright function is defined by the following expression: 
\begin{eqnarray}
W_{\lambda,\mu}\left(z\right)=\sum_{n=0}^{\infty}\frac{z^n}{n!\Gamma\left(\lambda n+\mu\right)}, \label{Wdef}
\end{eqnarray}
for every $\lambda >-1$, and $z,\mu \in \mathbb{C}$. In the present scenario, we consider uniquely positive values of the involved parameters, $\lambda,\mu>0$. The Wright GCSs fulfill the three required conditions. Particularly, this set of GCSs provides the following resolution of the identity:
\begin{eqnarray}
\int_{\mathbb{R}^2}U^{\left(W\right)}_{\lambda,\mu}\left(\left|z\right|^2\right)|z;\lambda,\mu \rangle\langle z;\lambda,\mu| d^2 z=I,
\label{completnessWCS}
\end{eqnarray}
 where $d^2 z=d\operatorname{Re}z  d\operatorname{Im}z$. The corresponding weight function $U^{\left(W\right)}_{\lambda,\mu}\left(u\right)$ is given by the following expression: 
\begin{eqnarray}
&&U^{\left(W\right)}_{\lambda,\mu}\left(u\right)=\pi^{-1}
W_{\lambda,\mu}\left(u\right) 
H^{2,0}_{0,2}\left[u\Bigg|\begin{array}{rr}\\ \left(0,1\right),\left(\mu-\lambda,\lambda\right)\end{array}\right], \hspace{1em}
\label{WeightWCS}
\end{eqnarray}
for every $u> 0$. The function $H^{2,0}_{0,2}\left[u\Bigg|\begin{array}{rr}\\ \left(0,1\right),\left(\mu-\lambda,\lambda\right)\end{array}\right]$ is a particular example of the Fox-$H$ function \cite{Foxdef,FoxHbook}. This special function is defined via a Mellin-Barnes type integral in the complex domain,
  \begin{eqnarray}
%&&\hspace{-6em}
&&H_{l,q}^{r,n}\left[z\Bigg|
\begin{array}{rr}
\left(o_1,O_1\right), \ldots,\left(o_l,O_l\right)\\
\left(v_1,V_1\right), \ldots,\left(v_q,V_q\right)\,\,
\end{array}
\right]=\frac{1}{2 \pi \imath} 
\int_{\mathcal{C}}
\frac{\Pi_{j=1}^r \Gamma\left(v_j+V_j s\right)
\Pi_{r=1}^n \Gamma\left(1-o_l-O_l s\right)
}{\Pi_{l=n+1}^l \Gamma\left(o_l+O_l s\right)
\Pi_{j=r+1}^q \Gamma\left(1-v_j-V_j s\right)
} \,z^{-s}  ds,\nonumber\\
&&\label{FoxHdef}
\end{eqnarray}
under the conditions that the poles of the Gamma functions which appear in the denominator of the integrand function, do not coincide. The empty products are interpreted as unity. The natural numbers $r,n,l,q$ fulfill the constraints: $0\leq r\leq q$, $0\leq n\leq l$, and $O_i, V_j\in \left(0,+\infty\right)$ for every $i=1,\cdots,l$ and $j=1,\cdots,q$. For the sake of shortness, refer to \cite{FoxHbook} for details on the contour path $\mathcal{C}$, the existence condition and related properties. Refer to \cite{Wcoher} for details 
on the Wright GCSs. Two Wright GCSs $|z_1;\lambda,\mu\rangle$ and $|z_2;\lambda,\mu\rangle$ are orthogonal to each others,
\begin{eqnarray}
&&\langle z_1;\lambda,\mu||z_2;\lambda,\mu\rangle= \left[W_{\lambda,\mu}\left(\left|z_1\right|^2\right)W_{\lambda,\mu}\left(\left|z_2\right|^2\right)\right]^{-1/2}
W_{\lambda,\mu}\left(z_1^{\ast}z_2\right)
=0, \label{orthogonalityWCSs}
\end{eqnarray}
in case the complex number $\left(z^{\ast}_1 z_2\right)$ is a zero of the Wright function, $W_{\lambda,\mu}\left(z_1^{\ast}z_2\right)=0$. The zeros of this function, studied in Ref.\cite{WzeroLucko}, form an infinite sequence. Again, two Wright GCSs are orthogonal to each others uniquely on a set of vanishing measure. For sufficiently large values of the magnitude, the zeros are simple and belong to the negative real axis. In these cases, the corresponding orthogonal Wright GCSs exhibit the following property: $\arg z_1-\arg z_2= \pm \pi$. The probability 
$p_{\lambda,\mu}\left(n, z\right)$ that the Wright GCS $|z;\alpha,\beta\rangle$ consists in $n$ excitations, i.e., the state $|n\rangle$, results to be
\begin{eqnarray}
p_{\lambda,\mu}\left(n, z\right)=\frac{\left|z\right|^{2n} }{W_{\lambda,\mu}\left(\left|z\right|^2\right)n!\Gamma\left(\lambda n + \mu\right) }, \label{pW}
\end{eqnarray}
for every $n \in \mathbb{N}$. This probability exhibits the following asymptotic behavior for high values of the number of excitations:
\begin{eqnarray}
&&p_{\lambda,\mu}\left(n, z\right)\sim
\frac{\left|z\right|^{2n} e^{\left(\lambda +1\right)n}
\lambda^{-\lambda n-\mu+\left(1/2\right)} n^{-\left(\lambda+1\right)n-\mu}   }{2 \pi W_{\lambda,\mu}\left(\left|z\right|^2\right) }
, \label{pWasympt}
\end{eqnarray}
as $n \gg 1$. 

 GCSs are considered in Ref. \cite{MLcoher} by substituting the term $n!$ with the arithmetic function $g(n)$ for every $n \in \mathbb{N}$, 
\begin{eqnarray}
|z;g\rangle= \left[\mathfrak{N}_g\left(|z|^2\right)\right]^{-1/2} \sum_{n=0}^{\infty}
\frac{z^n }{\sqrt{g\left(n\right)}}|n\rangle, \hspace{1em}\label{GCoherent1}
\end{eqnarray}
for every $z\in \mathbb{C}\setminus \left\{0\right\}$, while 
$|0;g\rangle=|0\rangle$. The arithmetic function $g(n)$ is required to fulfill the conditions which are reported below. The term $\mathfrak{N}_g\left(|z|^2\right)$ is the normalization factor,
\begin{eqnarray}
\mathfrak{N}_g\left(|z|^2\right)= \sum_{n=0}^{\infty}
\frac{|z|^{2n}}{g\left(n\right)}. \hspace{1em}\label{GNf}
\end{eqnarray}
Hence, the above series is requested to be convergent and positive for every $z\in \mathbb{C}\setminus \left\{0\right\}$,
\begin{eqnarray}
0< \sum_{n=0}^{\infty}
\frac{|z|^{2n}}{g\left(n\right)}<+\infty. \hspace{1em}\label{positivFinitenessGNf}
\end{eqnarray}
 The completeness property of the GCSs consists in the following resolution of the identity operator with a positive weight function:
\begin{eqnarray}
\int_{\mathbb{R}^2}U^{\left(G\right)}\left(\left|z\right|^2\right)|z \rangle\langle z| d^2 z=I,
\label{completnessGCS}
\end{eqnarray}
where $d^2 z=d\operatorname{Re}z  d\operatorname{Im}z$. The existence of GCSs, which are given by Eq. (\ref{GCoherent1}) and are created by the arithmetic function $g(n)$, is related to the Stieltjes power-moment problem \cite{AkhiezerMomentProb1965} and is discussed in Refs. \cite{Klauder1963,Klauder1995,MLcoherJMP,MLcoher}. Consider a general sequence $g(n)$ such that the constraint (\ref{positivFinitenessGNf}) holds for every $z\in \mathbb{C}\setminus \left\{0\right\}$; a function $U^{\left(G\right)}(u)$ which is positive for every $u > 0$, provides the required completeness condition (\ref{completnessGCS}) iff two series of determinants, related to the Hankel-Hadamard matrices, are positive. The choice of peculiar arithmetic functions $g(n)$ which substitute the factorial term $n!$ simplifies the search for the appropriate weight function $U^{\left(G\right)}(u)$. The inverse problem, i.e. finding positive weight functions $U^{\left(G\right)}(u)$ such that the corresponding arithmetic function $g(n)$ fulfills the condition (\ref{positivFinitenessGNf}), has also been studied. Refer to \cite{AkhiezerMomentProb1965,MLcoherJMP,MLcoher} for details.

The probability $p_{g}\left(n, z\right)$ that the GCS $|z;g\rangle$ consists in $n$ excitations is given by the form below,
\begin{eqnarray}
p_{g}\left(n, z\right)=\frac{\left|z\right|^{2n} }{\mathfrak{N}_g\left(|z|^2\right) g(n)}, \label{pgn}
\end{eqnarray}
for every $n \in \mathbb{N}$. For large values of the number of excitations, $n \gg 1$, this probability exhibits various behaviors which depend on the asymptotic properties of the sequence $g(n)$ as $n\to +\infty$. A variety of these behaviors can be described by introducing the auxiliary function $f(u)$. This function is required to be positive, $f(u)>0$, for every $u>0$ and to be locally integrable on $\left.\right]0,\infty\left[\right.$. Let a real constant $r_0$ exists such that the function $u^{r_0-1}f(u)$ is integrable in any finite interval $\left[0,u_1\right]$, for every $u_1>0$; let the following asymptotic behavior: $f(u)=O\left(u^{-r_1}\right)$ holds as $u\to+\infty$ for every $r_1>0$; let the function $f(u) $ behaves asymptotically as follows: $f\left(u\right)\sim \sum_{j=0}^{\infty}c_j e_{j} (u)$ for $u\to+\infty$, where $c_j$ are real-valued constants. The sequence of functions $e_{0} (u)$, $e_{1} (u), \ldots$, is requested to be asymptotic: $e_{j_2} (u)/e_{j_1} (u)\to 0$ as $u\to+\infty$, for every natural value of the indexes $j_1$ and $j_2$ such that $j_1<j_2$. Furthermore, the asymptotic sequences under study consist in combinations of stretched exponential decays, power laws and, possibly, natural powers of logarithmic forms,  
\begin{eqnarray}
e_j(u)=u^{-\nu_j} \exp\left(-w_j u^{\rho_j}\right)\ln^{l_j} u, \label{e2jv}
\end{eqnarray}
for every $j\in \mathbb{N}$, where $w_j>0$ and $l_j$ is a natural power, $l_j\in \mathbb{N}$. The involved parameters fulfill the following constraints: 
$0<\rho_{0} \leq \rho_{1}\leq \ldots$; $w_{j_1}<w_{j_2}$ if $\rho_{j_1}=\rho_{j_2}$; $\nu_{j_1}<\nu_{j_2}$ if $\rho_{j_1}=\rho_{j_2}$ and $w_{j_1}=w_{j_2}$; and $n_{j_1}>n_{j_2}$ if $\rho_{j_1}=\rho_{j_2}$, $w_{j_1}=w_{j_2}$ and $\nu_{j_1}=\nu_{j_2}$. It is also required that the Mellin transform $\hat{f}(s)$ exists for $\operatorname{Re} s\geq 1$. See Appendix for details on this integral transform. If the above conditions hold, the choice $g(n)=\hat{f}(n+1)$, for every $n \in \mathbb{N}$, realizes the condition (\ref{positivFinitenessGNf}) and the resolution of the identity (\ref{completnessGCS}) holds, with 
$U^{\left(G\right)}\left(u\right)=\mathfrak{N}_g\left(u\right)f(u)/\pi$ for every $u> 0$. Therefore, a legitimate class of GCSs is provided by Eq. (\ref{GCoherent1}) with $g(n)=\hat{f}(n+1)$, for every $n \in \mathbb{N}$, in case the function $f(u)$ fulfills the above-required conditions \cite{MellinAsymptSidi1985}. The corresponding probability $p_{g}\left(n, z\right)$, naturally given by the expression below,
\begin{eqnarray}
p_{g}\left(n, z\right)=\frac{\left|z\right|^{2n} }
{\mathfrak{N}_g\left(|z|^2\right) \hat{f}(n+1)}, \label{pgnf}
\end{eqnarray}
for every $n \in \mathbb{N}$, exhibits the following asymptotic behavior:
\begin{eqnarray}
&&\hspace{-1em}p_{g}\left(n, z\right)\sim \frac{
\rho_{j_0}^{l_{j_0}+1}\left|z\right|^{2n} w_{j_0}^{\rho_{j_0}^{-1}\left(n+1-\nu_{j_0}\right)}\exp\left(\rho_{j_0}^{-1} n\right)
\left(\rho_{j_0}^{-1} n\right)^{-\rho_{j_0}^{-1}
\left( n+1-\nu_{j_0}\right)+ \left(1/2\right)}
}{\left(2\pi\right)^{1/2}c_{j_0}\mathfrak{N}_g\left(|z|^2\right)
\ln^{l_{j_0}}\left(\rho_{j_0}^{-1} n\right)
}, \label{pgnfasympt}
\end{eqnarray}
for $n\gg 1$. The index $j_0$ is the smallest among the indexes $j^{\prime}$ such that the coefficient $c_{j^{\prime}}$ does not vanish, $c_{j^{\prime}}\neq 0$. According to Eq. (\ref{pgnfasympt}), the probability that the number of excitations is large, $n\gg 1$, is favored by the values $w_{j_0}>1$ with respect to the values $w_{j_0}<1$. Similarly, the presence of logarithmic powers, $l_{j_0}\geq 1$, disfavors the population of large number of excitations.

As an example of the above-defined class of GCSs, we consider the GCSs $|z;g_1\rangle$ which are generated by a special arithmetic function $g_1(n)$ reported below, 
\begin{eqnarray}
g_1(n)=\rho^{-1} w^{-\rho^{-1} \left(n +\nu+1\right)} \Gamma\left[\rho^{-1}\left(n+\nu+1\right)\right],
\label{gn1}
\end{eqnarray}
for every $n \in \mathbb{N}$, where $\nu \geq 0$, and $\rho,w>0$. The constraint (\ref{positivFinitenessGNf}) holds and the normalization factor
reads
\begin{eqnarray}
\mathfrak{N}_1\left(|z|^2\right)= \rho w^{\rho^{-1}\left(\nu+1\right)} E_{\rho^{-1},\rho^{-1}\left(\nu+1\right)}\left[w^{\rho^{-1}}\left|z\right|^2\right]. 
 \hspace{1em}\label{Ng1}
\end{eqnarray}
The probability $p_{1}\left(n, z\right)$ that the GCS $|z;g_1\rangle$  consists in $n$ excitations departs from the canonical Poisson distribution as below,
\begin{eqnarray}
p_{1}\left(n, z\right)=
\frac{w^{\rho^{-1}\left(n+\nu+1\right)} \left|z\right|^{2n}}{E_{\rho^{-1},\rho^{-1}\left(\nu+1\right)}\left[w^{\rho^{-1}}\left|z\right|^2\right] \Gamma\left[\rho^{-1}\left(n+\nu+1\right)\right]}, \label{pgn1}
\end{eqnarray}
for every $n \in \mathbb{N}$. For large values of the number of excitations, $n\gg 1$, this probability exhibits the asymptotic behavior which is given by the following form:
\begin{eqnarray}
&&p_{1}\left(n, z\right)\sim
\frac{w^{\rho^{-1}\left(n+\nu+1\right)} \left|z\right|^{2n} e^{\rho^{-1} n} \left(\rho^{-1} n\right)^{-\rho^{-1}\left(n+\nu+1\right)+\left(1/2\right)}}{\left(2\pi\right)^{1/2}E_{\rho^{-1},\rho^{-1}\left(\nu+1\right)}\left[w^{\rho^{-1}}\left|z\right|^{2n}\right]}, \label{pgn1asympt}
\end{eqnarray}
as $n \gg 1$.

Consider a general state ket $|\phi\rangle$ which belongs to the Hilbert space of the harmonic oscillator. The completeness property (\ref{completnessGCS}) of GCSs allows to express this state in terms of the GCSs:
\begin{eqnarray}
|\phi\rangle=
\int_{\mathbb{R}^2}U^{\left(G\right)}\left(\left|z\right|^2\right)\mathfrak{N}^{-1}_g\left(|z|^2\right)
\Phi\left(z^{\ast}\right)
|z;g \rangle d^2 z.
\label{PhiStatecompletnessGCS}
\end{eqnarray}
The complex-valued function $\Phi\left(z^{\ast}\right)$, defined as below,
\begin{eqnarray}
&&\Phi\left(z^{\ast}\right)
=\sum_{n=0}^{\infty}
\frac{\langle n || \phi \rangle }{\sqrt{g(n)}}\left( z^{\ast}\right)^n
, \label{phin}
\end{eqnarray}
is an analytic function of the complex variable $z^{\ast}$. This function is uniquely determined by the representation of the state $|\phi\rangle$ in the Fock basis. This type of functions forms an Hilbert space which is referred to as the Bargmann space. For the canonical CSs, the inner product is characterized by the measure $\left[\pi^{-1}\exp\left(-|z|^2\right)\right]$. For GCSs, the measure of the inner product is $U^{\left(G\right)}\left(\left|z\right|^2\right)\mathfrak{N}^{-1}_g\left(|z|^2\right)$. This expression is derived from the completeness relation (\ref{completnessGCS}) which provides the following form for the scalar product,
\begin{eqnarray}
\langle \psi||\phi\rangle=
\int_{\mathbb{R}^2}U^{\left(G\right)}\left(\left|z\right|^2\right)\mathfrak{N}^{-1}_g\left(|z|^2\right)
\Psi^{\ast}\left(z^{\ast}\right)
\Phi\left(z^{\ast}\right)
 d^2 z.
\label{innerproductPhiPsi}
\end{eqnarray}
The complex-valued function $\Psi\left(z^{\ast}\right)$, corresponding to the state $|\psi\rangle$,
\begin{eqnarray}
&&\Psi\left(z^{\ast}\right)
=\sum_{n=0}^{\infty}
\frac{\langle n || \psi \rangle }{\sqrt{g(n)}}\left( z^{\ast}\right)^n
, \label{psin}
\end{eqnarray}
is analytical of the variable $z^{\ast}$. Naturally, expressions (\ref{PhiStatecompletnessGCS})-(\ref{innerproductPhiPsi}) hold for Mittag-Leffler GCSs in case the arithmetic function $g(n)$ is substituted with the Gamma function, $\Gamma\left(\alpha n+ \beta\right)$, and the weight function $U^{\left(G\right)}\left(\left|z\right|^2\right)\mathfrak{N}^{-1}_g\left(|z|^2\right)$ with the expression $U^{\left(ML\right)}_{\alpha,\beta}\left(\left|z\right|^2\right)\left[E_{\alpha,\beta}\left(\left|z\right|^2\right)\right]^{-1}$. Similarly, expressions (\ref{PhiStatecompletnessGCS})-(\ref{innerproductPhiPsi}) hold for Wright GCSs in case the arithmetic function $g(n)$ is substituted with the expression $n! \Gamma\left(\lambda n+ \mu\right)$, and the weight function $U^{\left(G\right)}\left(\left|z\right|^2\right)\mathfrak{N}^{-1}_g\left(|z|^2\right)$ with the expression $U^{\left(W\right)}_{\lambda,\mu}\left(\left|z\right|^2\right)\left[W_{\lambda,\mu}\left(\left|z\right|^2\right)\right]^{-1}$.

\section{Truncated Mittag-Leffler, Wright and Generalized Coherent states }\label{3}

The definition of the CSs of an harmonic oscillator has been extended to the case where the corresponding Hilbert space $\mathcal{H}$ is finite-dimensional \cite{finiteDimH1,finiteDimH2,finiteDimH3,TCSsdefs1990,Truncosc1992,finiteDimCoher1,Truncosc21994,Truncosc1997,Truncosc2011,TruncoscXiv2014,Truncosc2020}. The truncated CSs are defined over the truncated Fock basis \cite{finiteDimCoher1,Truncosc2020} $\mathcal{F}_{k}\equiv \left\{|0\rangle,\cdots, |k\rangle\right\}$ as below,
\begin{eqnarray}
|z;k\rangle= \left[\exp_k \left(\left|z\right|^2\right)\right]^{-1/2} \sum_{n=0}^{k} \frac{z^n}{\sqrt{n!}}|n\rangle, \label{expCoherentTrunc1}
\end{eqnarray}
for every $z\in \mathbb{C}\setminus \left\{0\right\}$ and $k \in \mathbb{N}$, while $|0;k\rangle=|0\rangle$ for every $k \in \mathbb{N}$. The truncated exponential function \cite{TruncExp1,TruncExp2} is defined as follows: $\exp_k (z)\equiv\sum_{n=0}^k z^{n}/n!$, for every $z\in \mathbb{C}\setminus \left\{0\right\}$ and $k \in \mathbb{N}$, while $\exp_k (0)=1$ for every $k \in \mathbb{N}$. The truncated CSs exhibit the following resolution of the identity operator:
\begin{eqnarray}
\int_{\mathbb{R}^2}U^{\left(C,k\right)}\left(\left|z\right|^2\right)|z;k \rangle\langle z;k| d^2 z=I,
\label{completnessCk}
\end{eqnarray}
where $d^2 z=d\operatorname{Re}z  
d\operatorname{Im}z$, over the truncated Fock basis $\mathcal{F}_{k}$ \cite{finiteDimCoher1,Truncosc2020}. The weight function $U^{\left(C,k\right)}\left(u\right)$ reads as below,
\begin{eqnarray}
U^{\left(C,k\right)}\left(u\right)=\pi^{-1}\exp\left(-u\right)\exp_k\left(-u\right),
\label{weightCk}
\end{eqnarray}
for every $u \geq 0$. Two truncated CSs $|z_1;k\rangle$ and $|z_2;k\rangle$ are orthogonal to each others,
\begin{eqnarray}
&&\langle z_1;k||z_2;k\rangle= 
\left[\exp_k\left(\left|z_1\right|^2\right)
\exp_k\left(\left|z_2\right|^2\right)
\right]^{-1/2}%\nonumber \\ &&\times 
\exp_k\left(z_1^{\ast}z_2\right)
=0, \label{orthogonalityCCSsk}
\end{eqnarray}
in case the complex number $\left(z^{\ast}_1z_2\right)$ coincides with any of the $k$ zeros of the $k$-th order polynomial $\exp_k\left(z\right)$. The probability $p\left(n,k,z\right)$ that the truncated CS $|z;k\rangle$ consists in $n$ excitations, i.e., the state $|n\rangle$, is given by the expression below,
\begin{eqnarray}
p\left(n,k,z\right)=\frac{\left|z\right|^{2n}}{n! \exp_k\left(|z|^2\right) }, \label{pCCk}
\end{eqnarray}
for every $z\in \mathbb{C}\setminus \left\{0\right\}$, while $p\left(n,k,0\right)=\delta_{n,0}$ for every $n=0,\ldots,k$, where $\delta_{n,0}$ represents the Kronecker delta. Refer to \cite{finiteDimCoher1,Truncosc2020} for details.

The truncated Mittag-Leffler GCSs are defined from Eq. (\ref{MLCoherent1}) as a natural extension of the Mittag-Leffler GCSs,
\begin{eqnarray}
&|z;k;\alpha,\beta\rangle= \left[\mathfrak{E}_{k,\alpha,\beta}\left(\left|z\right|^2\right)\right]^{-1/2}
\sum_{n=0}^{k}
\frac{z^n }{\sqrt{\Gamma\left(\alpha n + \beta\right)}}|n\rangle,\label{MLTCkdef}
\end{eqnarray}
for every $z\in \mathbb{C}\setminus \left\{0\right\}$, $\alpha,\beta>0$, and $k \in \mathbb{N}$, while $|0;k;\alpha,\beta\rangle=|0\rangle$ for every $\alpha,\beta>0$ and $k \in \mathbb{N}$. The function $\mathfrak{E}_{k,\alpha,\beta}\left(z\right)$, appearing in the normalization constant, generalizes the truncated exponential function as follows: 
\begin{eqnarray}
\mathfrak{E}_{k,\alpha,\beta}\left(z\right)=
\sum_{n=0}^{k}
\frac{z^n }{\Gamma\left(\alpha n +\beta\right)}, \label{Ekdef}
\end{eqnarray}
for every $z\in \mathbb{C}\setminus \left\{0\right\}$, $\alpha,\beta>0$ and $k \in \mathbb{N}$; while $\mathfrak{E}_{k,\alpha,\beta}\left(0\right)=\left[\Gamma\left(\beta\right)\right]^{-1}$. The truncated exponential function is obtained for $\alpha=\beta=1$, i.e., $\exp_k\left(z\right)=\mathfrak{E}_{k,1,1}\left(z\right)$ for every $z \in \mathbb{C}$. Naturally, the Mittag-Leffler function is obtained as $k\to +\infty$, i.e., $\mathfrak{E}_{+\infty,\alpha,\beta}\left(z\right)=E_{\alpha,\beta}\left(z\right)$ for every allowed value of the involved parameters. Truncated Mittag-Leffler GCSs provide over the truncated Fock space $\mathcal{F}_{k}$ the following resolution of the identity with a positive weight function:
\begin{eqnarray}
\int_{\mathbb{R}^2}U^{\left(ML,k\right)}_{\alpha,\beta}\left(\left|z\right|^2\right)|z;k;\alpha,\beta \rangle\langle z;k;\alpha,\beta| d^2 z=I,
\label{completnesskMLCS}
\end{eqnarray}
 where $d^2 z=d\operatorname{Re}z  d\operatorname{Im}z$. The corresponding weight function $U^{\left(ML,k\right)}_{\alpha,\beta}\left(u\right)$ is given by the expression below, 
\begin{eqnarray}
U^{\left(ML,k\right)}_{\alpha,\beta}\left(u\right)=
\frac{u^{\left(\beta/\alpha\right)-1}\exp\left(-u^{1/\alpha}\right)\mathfrak{E}_{k,\alpha,\beta}\left(u\right)}{\pi \alpha },
\label{WeightTMLCSk}
\end{eqnarray}
for every $u> 0$. Two truncated Mittag-Leffler GCSs $|z_1;k;\alpha,\beta\rangle$ and $|z_2;k;\alpha,\beta\rangle$ are orthogonal to each others,
\begin{eqnarray}
&&\langle z_1;k;\alpha,\beta||z_2;k;\alpha,\beta\rangle= 
\left[\mathfrak{E}_{k,\alpha,\beta}\left(\left|z_1\right|^2\right)
\mathfrak{E}_{k,\alpha,\beta}\left(\left|z_2\right|^2\right)
\right]^{-1/2}
\mathfrak{E}_{k,\alpha,\beta}\left(z_1^{\ast}z_2\right)
=0, \label{productsCCSsk}
\end{eqnarray}
in case the complex number $\left(z^{\ast}_1z_2\right)$ coincides with any of the $k$ zeros of the $k$-th order polynomial $\mathfrak{E}_{k,\alpha,\beta}\left(z\right)$. The probability $p_{\alpha,\beta}\left(n,k,z\right)$ that the Mittag-Leffler GCS consists in $n$ excitations, i.e., the state $|n\rangle$, is
\begin{eqnarray}
p_{\alpha,\beta}\left(n,k,z\right)=\frac{\left|z\right|^{2n}
}{\Gamma\left(\alpha n +\beta\right)\mathfrak{E}_{k,\alpha,\beta}\left(\left|z\right|^2\right)}, \label{pMLk}
\end{eqnarray}
for every $z\in \mathbb{C}\setminus \left\{0\right\}$, $\alpha,\beta>0$ and $n=0,\ldots,k$, while $p_{\alpha,\beta}\left(n,k,0\right)=\delta_{n,0}$ for every $\alpha,\beta>0$ and $n=0,\ldots,k$. For large values of the number of excitations, $k\geq n\gg 1$, this probability is approximated with expression (\ref{pEasympt}) by substituting the function $E_{\alpha,\beta}\left(\left|z\right|^2\right)$ with $\mathfrak{E}_{k,\alpha,\beta}\left(\left|z\right|^2\right)$.

The truncated Wright GCSs are defined from Eq. (\ref{MLCoherent1}) as a natural extension of the Wright GCSs via the truncated Fock basis $\mathcal{F}_{k}$ as below,
\begin{eqnarray}
&&|z;k;\lambda,\mu\rangle= \left[\mathfrak{W}_{k,\lambda,\mu}\left(\left|z\right|^2\right)\right]^{-1/2} \sum_{n=0}^{k}
\frac{z^n }{\sqrt{n!\Gamma\left(\lambda n+\mu\right)}}|n\rangle, \label{WCoherentTruncdef}
\end{eqnarray}
for every $z\in \mathbb{C}\setminus \left\{0\right\}$, $\lambda,\mu>0$ and $k \in \mathbb{N}$; while $|0;k;\lambda,\mu\rangle=|0\rangle$ for every $\lambda,\mu>0$ and $k \in \mathbb{N}$. The function $\mathfrak{W}_{k,\lambda,\mu}\left(u\right)$, appearing in the normalization constant, is defined by the following form: 
\begin{eqnarray}
\mathfrak{W}_{k,\lambda,\mu}\left(z\right)=
\sum_{n=0}^{k}
\frac{z^n }{n!\Gamma\left(\lambda n +\mu\right)}, \label{Wkdef}
\end{eqnarray}
for every $z\in \mathbb{C}\setminus \left\{0\right\}$, $\lambda,\mu>0$ and $k \in \mathbb{N}$, while $\mathfrak{W}_{k,\lambda,\mu}\left(0\right)=\left[\Gamma\left(\mu\right)\right]^{-1}$, for every $\lambda,\mu>0$, and $k \in \mathbb{N}$. This function produces the Wright function as $k\to +\infty$, i.e., $\mathfrak{W}_{+\infty,\lambda,\mu}\left(z\right)=W_{\lambda,\mu}\left(z\right) $ for every $z\in \mathbb{C}$ and $\lambda,\mu>0$. Truncated Wright GCSs provide over the truncated Fock space $\mathcal{F}_{k}$ the following resolution of the identity:
\begin{eqnarray}
\int_{\mathbb{R}^2}U^{\left(W,k\right)}_{\lambda,\mu}\left(\left|z\right|^2\right)|z \rangle\langle z| d^2 z=I,
\label{completnesskWCS}
\end{eqnarray}
 where $d^2 z=d\operatorname{Re}z  d\operatorname{Im}z$. The corresponding weight function $U^{\left(W,k\right)}_{\lambda,\mu}\left(u\right)$ is given by the expression below, 
\begin{eqnarray}
&&U^{\left(W,k\right)}_{\lambda,\mu}\left(u\right)=\pi^{-1}
\mathfrak{W}_{k,\lambda,\mu}\left(u\right)
%\nonumber \\&&\times 
H^{2,0}_{0,2}\left[u\Bigg|\begin{array}{rr}\\ \left(0,1\right)\left(\mu-\lambda,\lambda\right)\end{array}\right], \hspace{1em}
\label{WeightTrWCS}
\end{eqnarray}
for every $u>0$. Two truncated Wright GCSs $|z_1;k;\lambda,\mu\rangle$ and $|z_2;k;\lambda,\mu\rangle$ are orthogonal to each others,
\begin{eqnarray}
&&\langle z_1;k;\lambda,\mu||z_2;k;\lambda,\mu\rangle= 
\left[\mathfrak{W}_{k,\lambda,\mu}\left(\left|z_1\right|^2\right)
\mathfrak{W}_{k,\lambda,\mu}\left(\left|z_2\right|^2\right)
\right]^{-1/2}%\nonumber \\ &&\times 
\mathfrak{W}_{k,\lambda,\mu}\left(z_1^{\ast}z_2\right)
=0, \label{productsWCSsk}
\end{eqnarray}
in case the complex number $\left(z^{\ast}_1 z_2\right)$ coincides with any of the $k$ zeros of the $k$-th order polynomial $\mathfrak{W}_{k,\lambda,\mu}\left(z\right)$. The probability $p_{\lambda,\mu}\left(n,k,z\right)$ that the truncated Wright GCS $|z;k;\lambda,\mu\rangle$ consists in $n$ excitations, i.e., the state $|n\rangle$, is
\begin{eqnarray}
p_{\lambda,\mu}\left(n,k,z\right)=\frac{\left|z\right|^{2n}
}{n!\Gamma\left(\lambda n+ \mu\right)\mathfrak{W}_{k,\lambda,\mu}\left(\left|z\right|^2\right)}, \label{pWk}
\end{eqnarray}
for every $z\in \mathbb{C}\setminus \left\{0\right\}$, $\lambda,\mu>0$ and $n=0,\ldots,k$, while $p_{\lambda,\mu}\left(n,k,0\right)=\delta_{n,0}$ for every $\lambda,\mu>0$ and $n=0,\ldots,k$. For high values of the number of excitations, $k\geq n\gg 1$, this probability is approximated by expression (\ref{pWasympt}) if the function $W_{\lambda,\mu}\left(\left|z\right|^2\right)$ is substituted with the function $\mathfrak{W}_{k,\lambda,\mu}\left(\left|z\right|^2\right)$.

Truncated GCSs are defined from Eq. (\ref{GCoherent1}) as a natural extension of the GCSs over the truncated Fock state $\mathcal{F}_{k}$ as below, 
\begin{eqnarray}
|z;k;g\rangle= \left[\mathfrak{N}_{k,g}\left(|z|^2\right)\right]^{-1/2} \sum_{n=0}^{k}
\frac{z^n }{\sqrt{g\left(n\right)}}|n\rangle, \label{GCoherent1k}
\end{eqnarray}
for every $z\in \mathbb{C}\setminus \left\{0\right\}$ 
and $k \in \mathbb{N}$, while $|0;k,g\rangle=|0\rangle$ 
for every arithmetic function $g(n)$ fulfilling the above-required condition and $k \in \mathbb{N}$. The term $\mathfrak{N}_{k,g}\left(|z|^2\right)$ appearing in the normalization factor reads as below, 
\begin{eqnarray}
\mathfrak{N}_{k,g}\left(|z|^2\right)= \sum_{n=0}^{k}
\frac{|z|^{2n}}{g\left(n\right)}. \hspace{1em}\label{GNfk}
\end{eqnarray}
Differently from the case of GCSs, constraint (\ref{positivFinitenessGNf}) is not required to hold for truncated GCSs as the maximum value $k$ of the index $n$ is finite. Truncated GCSs provide over the truncated Fock basis $\mathcal{F}_{k}$ the following resolution of the identity with a positive weight function,
\begin{eqnarray}
\int_{\mathbb{R}^2}U^{\left(G,k\right)}\left(\left|z\right|^2\right)|z;k;g \rangle\langle z;k;g| d^2 z=I,
\label{completnessGCSk}
\end{eqnarray}
where $d^2 z=d \operatorname{Re} z  d \operatorname{Im} z$. 
Again, the weight function $U^{\left(G,k\right)(u)}$ can be determined from the positive sequence $g(0),\ldots , g(k)$, in case the Mellin transform $\hat{f}(s)$ of the positive auxiliary function $f(u)$ exists for $1\leq \operatorname{Re} s \leq k+1$ and fulfills the relation $\hat{f}(n+1)=g(n)$ for every $n=0,\ldots,k$. Under these conditions, the weight functions reads $U^{\left(G,k\right)}\left(u\right)=\mathfrak{N}_{k,g}\left(u\right)f(u)/\pi$ for every $u> 0$. Consider a general state ket $|\phi^{\prime}\rangle$ which belongs to the finite-dimensional Hilbert space with Fock basis $\mathcal{F}_{k}$. The completeness property (\ref{completnessGCSk}) allows to express this state in terms of the truncated GCSs,
\begin{eqnarray}
|\phi^{\prime}\rangle=
\int_{\mathbb{R}^2}U^{\left(G,k\right)}\left(\left|z\right|^2\right)\mathfrak{N}^{-1}_{k,g}\left(|z|^2\right)
\Phi_k\left(z^{\ast}\right)
|z;k;g \rangle d^2 z,
\label{PhiStatecompletnessGCSk}
\end{eqnarray}
where $\Phi_k\left(z^{\ast}\right)$ is a polynomial of maximum degree $k$ in the complex variable $z^{\ast}$,
\begin{eqnarray}
&&\Phi_k\left(z^{\ast}\right)
=\sum_{n=0}^{k}
\frac{\langle n || \phi^{\prime} \rangle }{\sqrt{g(n)}}\left( z^{\ast}\right)^n. \label{phink}
\end{eqnarray}
This polynomial is uniquely determined by the representation of the state $|\phi^{\prime}\rangle$ in the finite-dimensional Fock basis $\mathcal{F}_{k}$. For truncated GCSs, the measure of the inner product is given by the following form: $U^{\left(G,k\right)}\left(\left|z\right|^2\right)\mathfrak{N}^{-1}_{k,g}\left(|z|^2\right)$. In fact, we find 
\begin{eqnarray}
\langle \psi^{\prime}||\phi^{\prime}\rangle=\int_{\mathbb{R}^2}U^{\left(G,k\right)}\left(\left|z\right|^2\right)\mathfrak{N}^{-1}_{k,g}\left(|z|^2\right)\Psi_k^{\ast}\left(z^{\ast}\right)\Phi_k\left(z^{\ast}\right) d^2 z.
\label{innerproductPhiPsik}
\end{eqnarray}
The ket $|\psi^{\prime}\rangle$ is a general state of the finite-dimensional Fock space $\mathcal{F}_{k}$. The function $\Psi_k\left(z^{\ast}\right)$ is uniquely determined by the representation of the state $|\phi^{\prime}\rangle$ in the finite-dimensional Fock space $\mathcal{F}_{k}$,
\begin{eqnarray}
&&\Psi_k\left(z^{\ast}\right)
=\sum_{n=0}^{k}
\frac{\langle n || \psi^{\prime} \rangle }{\sqrt{g(n)}}\left( z^{\ast}\right)^n
, \label{psink}
\end{eqnarray}
and is analytical in the variable $z^{\ast}$. Expression (\ref{innerproductPhiPsik}) is derived from the completeness relation (\ref{completnessGCSk}). Naturally, expressions (\ref{PhiStatecompletnessGCS})-(\ref{innerproductPhiPsi}) hold for the truncated Mittag-Leffler GCSs in case the sequence $g(n)$ is substituted with $\Gamma\left(\alpha n+ \beta\right)$, and the weight function $U^{\left(G\right)}\left(\left|z\right|^2\right)\mathfrak{N}^{-1}_g\left(|z|^2\right)$ with the expression $U^{\left(ML\right)}_{\alpha,\beta}\left(\left|z\right|^2\right)\left[E_{\alpha,\beta}\left(\left|z\right|^2\right)\right]^{-1}$. Similarly,  expressions (\ref{PhiStatecompletnessGCS}) and (\ref{innerproductPhiPsi}) hold for the truncated Wright GCSs in case the arithmetic function $g(n)$ is substituted with the form $n! \Gamma\left(\lambda n+ \mu\right)$, and the function $U^{\left(G\right)}\left(\left|z\right|^2\right)\mathfrak{N}^{-1}_g\left(|z|^2\right)$ with the following expression: \\ $U^{\left(W\right)}_{\lambda,\mu}\left(\left|z\right|^2\right)\left[W_{\lambda,\mu}\left(\left|z\right|^2\right)\right]^{-1}$.

The probability $p_{g}\left(n,k,z\right)$ that the truncated GCS $|z;k;g \rangle$ consists in $n$ excitations, i.e., the state $|n\rangle$, results to be
\begin{eqnarray}
p_{g}\left(n,k,z\right)=\frac{\left|z\right|^{2n} }
{\mathfrak{N}_{k,g}\left(|z|^2\right) \hat{f}(n+1)}, \label{pgnfk}
\end{eqnarray}
for every $z\in \mathbb{C}\setminus \left\{0\right\}$ and $k \in \mathbb{N}$, while $p_{g}\left(n,k,0\right)=\delta_{n,0}$, for every $n=0,\ldots,k$, and $k \in \mathbb{N}$. If the auxiliary function $f(u)$ fulfills the additional properties which are required for the GCSs in the third last paragraph of the previous Section, the probability $p_{g}\left(n,k,z\right)$ is properly approximated by expression (\ref{pgnfasympt}) for large numbers of excitations, $k\geq n\gg 1$. The factor $\mathfrak{N}_{g}\left(|z|^2\right)$ must be substituted with the term $\mathfrak{N}_{k,g}\left(|z|^2\right)$. For small values of the label, the probability $p_{g}\left(n,k,z\right)$ behaves as follows:
\begin{eqnarray}
p_{g}\left(n,k,z\right)\sim \frac{ g(0)}{g(n)}\left|z\right|^{2n}\left(1-\frac{ g(0)}{g(1)}\left|z\right|^{2}\right), \label{pgnfkz0}
\end{eqnarray}
as $\left|z\right|\to 0^+$, with $z\neq 0$, for every $n=0,\ldots,k$, and $k=1,2\ldots$. Particularly, we find 
\begin{eqnarray}
&&p_{g}\left(0,k,z\right)\sim 1-\frac{ g(0)}{g(1)}\left|z\right|^{2}, \label{pg0fkz0}\\
&&p_{g}\left(k,k,z\right)\sim \frac{ g(0)}{g(k)}\left|z\right|^{2k}\left(1-\frac{ g(0)}{g(1)}\left|z\right|^{2}\right). \label{pgkfkz0}
\end{eqnarray}
Hence, at small nonvanishing values of the label the probability $p_{g}\left(n,k,z\right)$ tends to unity for $n=0$ and vanishes for every $n=1,\ldots,k$. Instead, for large values of the label, the probability $p_{g}\left(n,k,z\right)$ behaves as below,
\begin{eqnarray}
p_{g}\left(n,k,z\right)\sim \frac{ g(k)}{g(n)}\left|z\right|^{2(n-k)}\left(1-\frac{ g(k)}{g(k-1)}\left|z\right|^{-2}\right), \label{pgnfkzinfty}
\end{eqnarray}
as $\left|z\right|\to +\infty$, for every  $k=1,2\ldots$, and, particularly,
\begin{eqnarray}
&&p_{g}\left(0,k,z\right)\sim \frac{ g(k)}{g(0)}\left|z\right|^{-2k}\left(1-\frac{ g(k)}{g(k-1)}\left|z\right|^{-2}\right), \label{pg0fkzinfty}\\
&&p_{g}\left(k,k,z\right)\sim 1-\frac{ g(k)}{g(k-1)}\left|z\right|^{-2}. \label{pgkfkzinfty}
\end{eqnarray}
Therefore, at large values of the label, the probability $p_{g}\left(n,k,z\right)$ vanishes for every $n=0,\ldots,k-1$, and tends to unity for $n=k$. The asymptotic forms (\ref{pgnfkz0})-(\ref{pgkfkzinfty}) provide the asymptotic behaviors of the probability $p_{\alpha,\beta}\left(n,k,z\right)$, for truncated Mittag-Leffler GCSs, or $p_{\lambda,\mu}\left(n,k,z\right)$, for truncated Wright GCSs, as $\left|z\right|\to 0^+$ and $\left|z\right|\to +\infty$, in case $g(n)=\Gamma\left(\alpha n+\beta\right)$ or $g(n)=n!\Gamma\left(\lambda n+\mu\right)$, respectively, for every $n=0,\ldots,k$, and $\alpha,\beta,\lambda,\mu>0$.

\begin{figure}
\includegraphics{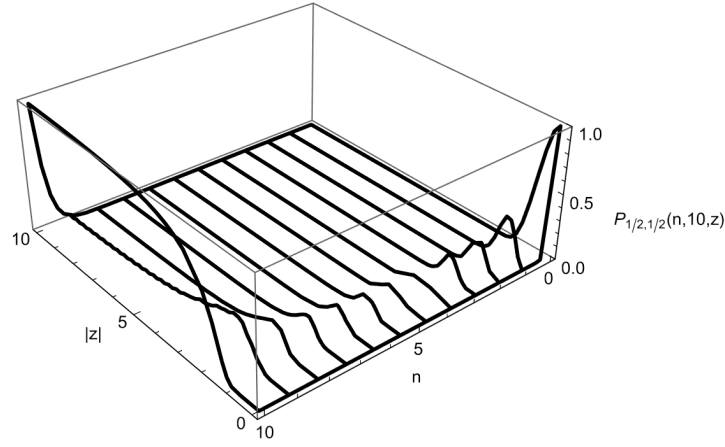}
\caption{(Color online) The probability $p_{\alpha,\beta}\left(n,k,z\right)$ for $\alpha=\beta=1/2$, $k=10$, $0\leq \left|z\right|\leq 10$, and $n=0,\ldots,10$.\label{Fig1}}
\end{figure}

\begin{figure}
\includegraphics{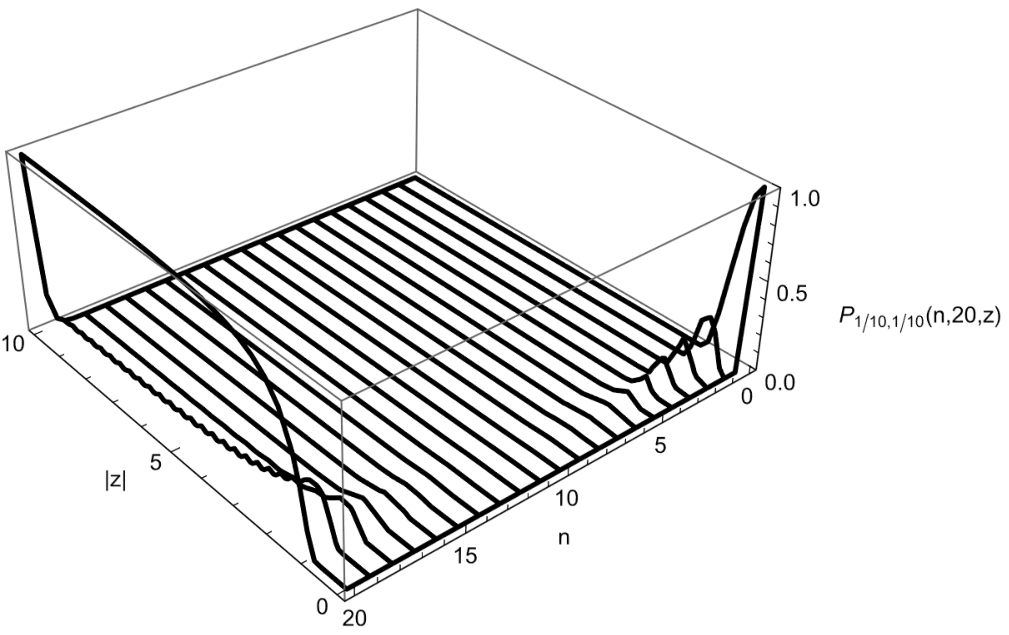}
\caption{(Color online) The probability $p_{\alpha,\beta}\left(n,k,z\right)$ for $\alpha=\beta=1/10$, $k=20$, $0\leq \left|z\right|\leq 10$ and $n=0,\ldots,20$.\label{Fig2}}
\end{figure}

\begin{figure}
\includegraphics{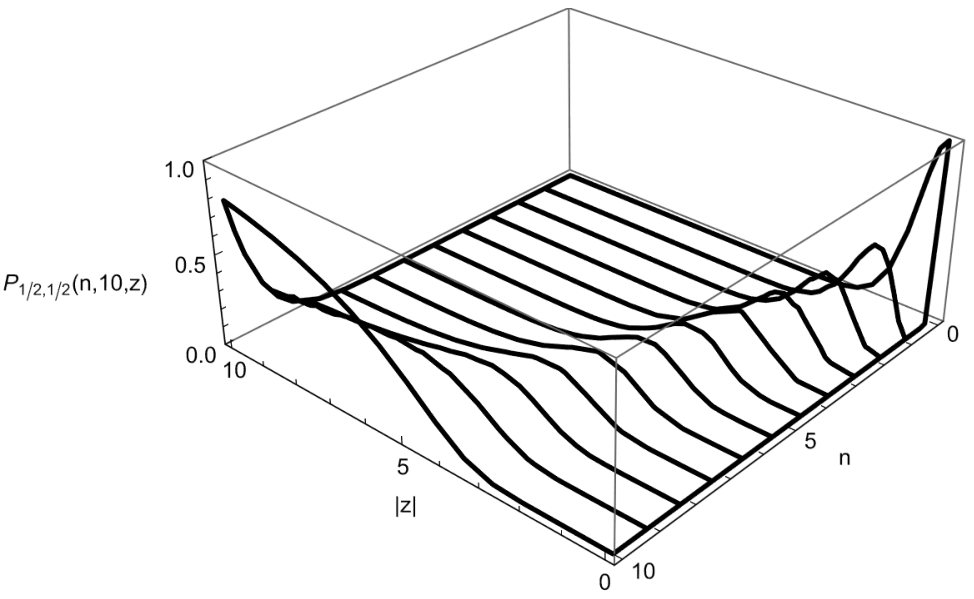}
\caption{(Color online) The probability $p_{\lambda,\mu}\left(n,k, z\right)$ for $k=10$, $\lambda=\mu=1/2$,
 $0\leq \left|z\right|\leq 10$ and $n=0,\ldots,10$.
\label{Fig3}}
\end{figure}

\begin{figure}
\includegraphics{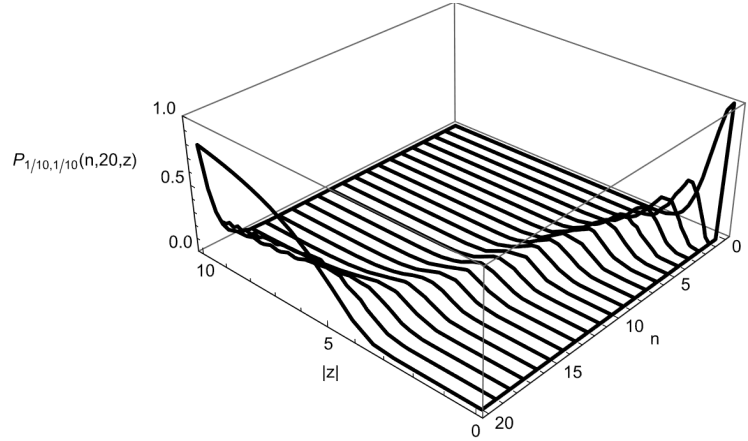}
\caption{(Color online) The probability $p_{\lambda,\mu}\left(n,k, z\right)$ for $k=20$, $\lambda=\mu=1/10$,
 $0\leq \left|z\right|\leq 10$ and $n=0,\ldots,20$.
\label{Fig4}}
\end{figure}

For the truncated Mittag-Leffler GCSs, the probability $p_{\alpha,\beta}\left(n,k,z\right)$ is displayed in Figures \ref{Fig1} and \ref{Fig2} for $k=10$ and 
$k=20$, respectively, and $0\leq \left|z\right|\leq 10$. For the truncated Wright GCSs, the probability $p_{\lambda,\mu}\left(n,k,z\right)$ is displayed in Figures \ref{Fig3} and \ref{Fig4} for $k=10$ and $k=20$, respectively, and $0\leq \left|z\right|\leq 10$. In every Figure, the corresponding probability exhibits the maximal value, unity, for vanishing value of the label, $z=0$, and $n=0$. Also, the maximum value is approached for large values of the label and $n=k$. Otherwise, the probability vanishes for vanishing value of the label, $z=0$ and $n=1,\ldots,k$, and tends to vanish for large values of the label. These behaviors are in accordance with the theoretical results analysis.

\section{The Mandel $Q$ parameter
}\label{4}

If the quantum system is described by a canonical CS the distribution of the number of excitations is Poissonian. The Mandel $Q$ parameter estimates the deviation from the Poisson statistics \cite{Mandel1979}. In fact, this parameter is defined via the number operator $\hat{N}$ as follows:
\begin{eqnarray}
Q=\frac{\langle \hat{N}^2 \rangle -\langle \hat{N} \rangle^2}
{\langle \hat{N} \rangle}-1. \label{Qdef}
\end{eqnarray}
The terms $\langle \hat{N}^2 \rangle$ and $\langle \hat{N} \rangle$ are the expectation values of the operators $\hat{N}^2$ and $\hat{N}$, respectively. Positive values of the Mandel parameter, $Q>0$, correspond to super-Poissonian distributions of the number of excitations. In this case the variance is larger than the mean value of the number of excitations, $\left[\langle \hat{N}^2 \rangle-\langle \hat{N} \rangle^2\right]>\langle \hat{N} \rangle$. The vanishing value, $Q=0$, corresponds to the Poisson distribution of the number of excitations. In this case the variance coincides with the mean value of the number of excitation, 
$\left[\langle \hat{N}^2 \rangle-\langle \hat{N} \rangle^2\right]=\langle \hat{N} \rangle$. Negative values of the Mandel parameter, $Q<0$, correspond to sub-Poissonian distributions of the number of excitations. In this case the variance is smaller than the mean value of the number of excitations, $\left[\langle \hat{N}^2 \rangle-\langle \hat{N} \rangle^2\right]<\langle \hat{N} \rangle$. This condition indicates the non-classical nature of the system in case the dimension is infinite, $k=+\infty$.  

If the Hilbert space of the quantum system is finite-dimensional, $\left(k+1\right)$, with $k<+\infty$, the probability $p_{g}\left(n,k,z\right)$, that a truncated GCS $|z;k;g \rangle$ consists in $n$ excitations, with $0\leq n \leq k$, produces the form $p\left(n,k,z\right)$, given by Eq. (\ref{pCCk}), in case the arithmetic function $g(n)$ coincides with the factorial term, $\Gamma\left(n+1\right)$. The probability $p\left(n,k,z\right)$ produces the canonical Poisson distribution in the natural variable $n$ uniquely for $k = +\infty$. However, the sign of the Mandel $Q$ parameter reflects the above-described interplay between variance and mean number of excitations also in the finite-dimensional case, $k<+\infty$. The Mandel $Q$ parameter and the sign are studied in details in Ref. \cite{Truncosc2020} for truncated CSs. Here, we evaluate the behavior and the sign of the Mandel $Q$ parameter of GCSs and truncated GCSs, for small nonvanishing values of the label, and truncated GCSs, for large values of the label.

Consider the general expression (\ref{GCoherent1k}) of the truncated GCSs. By definition, the arithmetic function $g(n)$ fulfills the required condition of positivity, $g(n)>0$, and completeness, Eq. (\ref{completnessGCSk}) with the positive weight function $U^{\left(G,k\right)}\left(u\right)$, for every $n=0,\ldots,k$, and for every $k\in \mathbb{N}$. For $k=0$, the truncated GCS coincides with the vacuum state $|0\rangle$. For $k=1$, the truncated GCS represents a two-level system,
\begin{eqnarray}
|z;1;g\rangle= \left[\mathfrak{N}_{1,g}\left(|z|^2\right)\right]^{-1/2} \left(\frac{1}{\sqrt{g(0)}}|0\rangle
+ \frac{z}{\sqrt{g(1)}}|1\rangle\right). \hspace{1em}\label{GCoherent11}
\end{eqnarray}
The term $\mathfrak{N}_{1,g}\left(|z|^2\right)$ is the normalization factor, 
\begin{eqnarray}
\mathfrak{N}_{1,g}\left(|z|^2\right)=  \left(\frac{1}{g(0)}
+ \frac{|z|^2}{g(1)}\right).\label{GNf1}
\end{eqnarray}
We are interested to the statistics which describes the numbers of excitations for high-dimensional Fock spaces. However, for the sake of completeness, we consider also low-dimensional Fock spaces. For $k=1$, the Mandel parameter, given by the expression below for every $z\in \mathbb{C}\setminus \left\{0\right\}$, 
\begin{eqnarray}
&&Q_1\left(\left|z\right|^2\right)=-
\frac{g(0)\left|z\right|^2}
{g(1)+g(0)\left|z\right|^{2}},  \label{QTGCSk1s}
\end{eqnarray}
is negative, $Q_1\left(\left|z\right|^2\right)<0$.

For $k=2$, the Mandel parameter $Q_2\left(\left|z\right|^2\right)$ reads as follows for every $z\in \mathbb{C}\setminus \left\{0\right\}$,
\begin{eqnarray}
&&Q_2\left(\left|z\right|^2\right)=\left|z\right|^2
\left\{ \frac{2 g(1)}{g(2)+2 g(1) \left|z\right|^2}
-\frac{ g(0)\left[g(2)+2 g(1) \left|z\right|^2\right]}{
g(1)g(2)+g(0)g(2)\left|z\right|^2 + g(0)g(1)\left|z\right|^4}
\right\}. \nonumber \\&& \label{Q2TGCSs}
\end{eqnarray}
 This parameter is positive, $Q_2\left(\left|z\right|^2\right)>0$,
if $g(0)g(2)/g^2(1)<2$ for $0<\left|z\right|
<\zeta_0$, where
$$\zeta_0=\sqrt{\frac{g(2)}{2g(1)}\left(\sqrt{\frac{4g^2(1)}{g(0)g(2)}-1}-1\right)}.$$
The Mandel parameter $Q_2\left(\left|z\right|^2\right)$ vanishes, $Q_2\left(\left|z\right|^2\right)=0$, if $g(0)g(2)/g^2(1)<2$ for $\left|z\right|
=\zeta_0$. This parameter is negative, $Q_2\left(\left|z\right|^2\right)<0$, if 
$g(0)g(2)/g^2(1)\geq 2$ for every $z\in \mathbb{C}\setminus \left\{0\right\}$, or if $g(0)g(2)/g^2(1)<2$ for $\left|z\right|>\zeta_0$.

For $k\geq 2$, the Mandel parameter $Q_k\left(\left|z\right|^2\right)$ is given by the expression below,
\begin{eqnarray}
&&Q_k\left(\left|z\right|^2\right)=\left|z\right|^2
\left[\frac{\sum_{n=0}^{k-2}(n+1)(n+2)\left|z\right|^{2n}/g(n+2)}
{\sum_{n=0}^{k-1}(n+1)\left|z\right|^{2n}/g(n+1)}-\frac{\sum_{n=0}^{k-1}(n+1)\left|z\right|^{2n}/g(n+1)}
{\sum_{n=0}^{k}\left|z\right|^{2n}/g(n)}\right], \nonumber \\&& \label{QTGCSs}
\end{eqnarray}
for every $z\in \mathbb{C}\setminus \left\{0\right\}$. The above expression shows that for every $k=2,3,\ldots,+\infty$, the Mandel parameter is positive (negative), $Q_k\left(\left|z\right|^2\right)>_{\left(<\right)}0$ as $\left|z\right|\to 0^+$, with $z \neq 0$, in case the following constraint holds:
\begin{eqnarray}
\frac{g(0)g(2)}{g^2(1)}<_{\left(>\right)}2.  \label{condg0g1g2QMm0}
\end{eqnarray}
Therefore, for small nonvanishing values of the label, the GCSs exhibit super-Poissonian (sub-Poissonian) statistics of the number of excitations in case condition (\ref{condg0g1g2QMm0}) holds.

The sign of the Mandel parameter depends on the value of the natural number $k$, and, therefore, on the dimension of the truncated Fock space, in case 
\begin{eqnarray}
\frac{g(0)g(2)}{g^2(1)}=2.  \label{condg0g1g2Qeq0}
\end{eqnarray}
In fact, for small nonvanishing values of the label, $\left|z\right|\to 0^+$ and $z \neq 0$, we find $Q_2\left(\left|z\right|^2\right)<0$ if condition (\ref{condg0g1g2Qeq0}) holds; $Q_3\left(\left|z\right|^2\right)>_{\left(<\right)}0$ if 
\begin{eqnarray}
\frac{g(0)g(3)}{g(1)g(2)}<_{\left(>\right)}3;  \label{condg0g3g1g2Qeq0}
\end{eqnarray}
and $Q_3\left(\left|z\right|^2\right)<0$ if
\begin{eqnarray}
\frac{g(0)g(3)}{g(1)g(2)}=3.  \label{condg0g3g1g2Qeq01}
\end{eqnarray}
For $k\geq 4$ and small nonvanishing values of the label, the sign of the Mandel parameter is determined by further conditions which are obtained via higher order approximations.

For large values of the label, $\left|z\right|\to +\infty$, and for every $k= 2,3,\ldots$, the Mandel parameter tends to the opposite of unity as below,
\begin{eqnarray}
&&\hspace{-1.2em}Q_{k}\left(\left|z\right|^2\right)
\sim -1+ \frac{g(k)}{k g(k-1)}|z|^{-2}. \label{Qkinfty}
 \end{eqnarray}

\subsection{Special cases}\label{41}
At this stage, we consider the special cases of the 
GCSs and truncated  GCSs which consist in the Mittag-Leffler GCSs and truncated Mittag-Leffler GCSs, Eq. (\ref{MLTCkdef}), or the Wright GCSs and truncated Wright GCSs, Eq. (\ref{WCoherentTruncdef}). For Mittag-Leffler GCSs and truncated Mittag-Leffler GCSs the Mandel parameter $Q_k\left(\left|z\right|^2\right)$ is positive (negative) for every $k= 2,3,\ldots,+\infty$, for small nonvanishing values of the label, $\left|z\right|\to 0^+ $, with $z\neq 0$, and for values of the parameters $\alpha$ and $\beta$ such that the following relation holds:
$$\frac{\Gamma\left(\beta\right)\Gamma\left(2\alpha+\beta\right)}{
\Gamma^2\left(\alpha+\beta\right)}<_{\left(>\right)}2.$$ Instead, for Wright GCSs and truncated Wright GCSs the Mandel parameter $Q_k\left(\left|z\right|^2\right)$ is uniquely negative for every $k= 2,3,\ldots,+\infty$, small nonvanishing values of the label, $\left|z\right|\to 0^+ $, with $z\neq 0$, as the following inequality about the ratio of Gamma functions holds \cite{Gammaineq}:
$$\frac{\Gamma\left(\mu\right)\Gamma\left(2\lambda+\mu\right)
}{\Gamma^2\left(\lambda+\mu\right)}>1,$$
for every $\lambda,\mu>0$. The Mandel parameter 
$Q_k\left(\left|z\right|^2\right)$ is uniquely negative for both truncated Mittag-Leffler GCSs and truncated Wright GCSs 
as $\left|z\right|\to +\infty$, for every $k=2,3,\ldots$ and every $\alpha,\beta,\lambda,\mu>0$.

\begin{figure}
\includegraphics{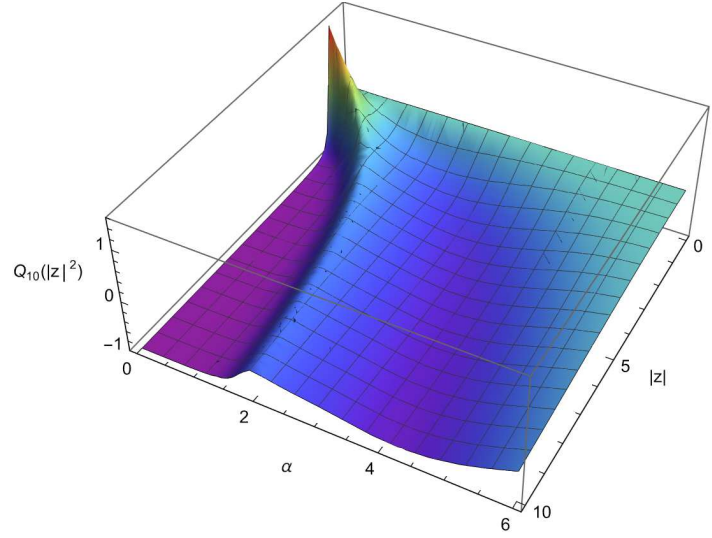}
\caption{(Color online) The Mandel parameter $Q_k\left(\left|z\right|^2\right)$ for truncated Mittag-Leffler GCSs in case $k=10$, $\beta=1/2$, $0<\alpha\leq 6$, and $0< \left|z\right|\leq 10$.
\label{Fig5}}
\end{figure}

\begin{figure}
\includegraphics{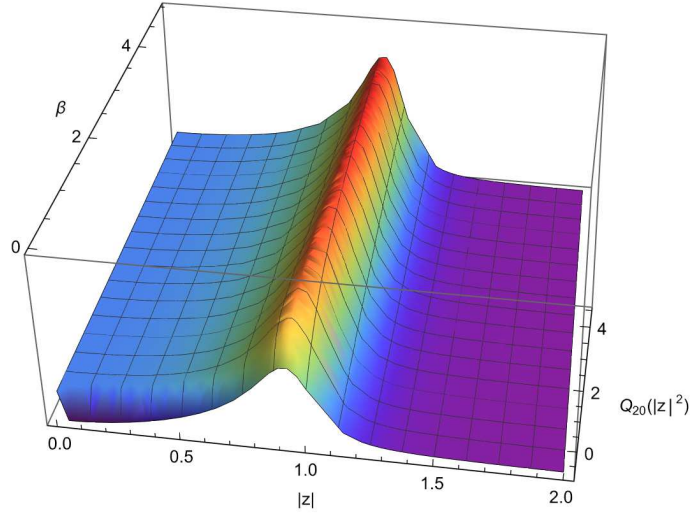}
\caption{(Color online) The Mandel parameter $Q_k\left(\left|z\right|^2\right)$ for truncated Mittag-Leffler GCSs in case $k=20$, $\alpha=1/10$, $0<\beta\leq 6$, and $0< \left|z\right|\leq 2$.
\label{Fig6}}
\end{figure}

\begin{figure}
\includegraphics{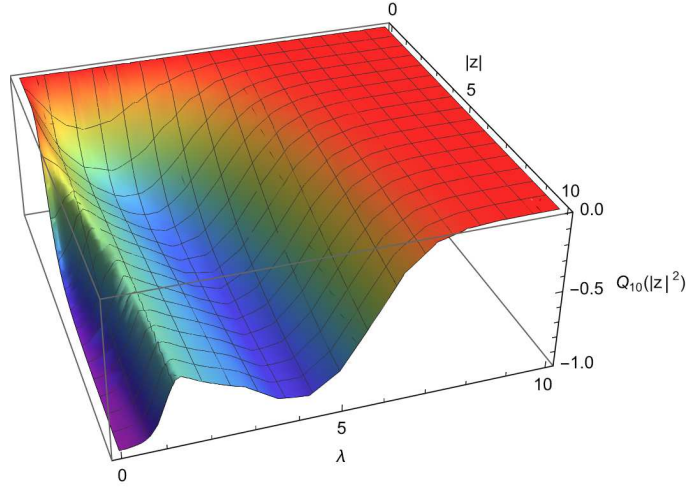}
\caption{(Color online) The Mandel parameter $Q_k\left(\left|z\right|^2\right)$ for truncated Wright GCSs with $k=10$, $\mu=1/2$, $0<\lambda\leq 6$, and $0< \left|z\right|\leq 10$.
\label{Fig7}}
\end{figure}

\begin{figure}
\includegraphics{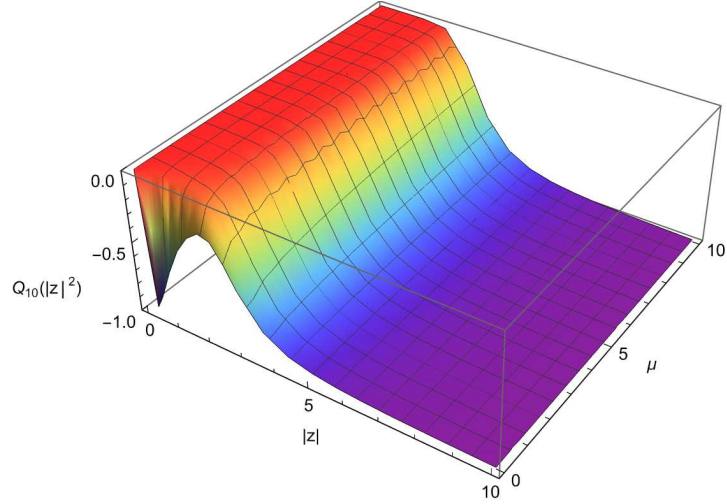}
\caption{(Color online) The Mandel parameter $Q_k\left(\left|z\right|^2\right)$ for truncated WCSs with $k=10$, $\lambda=1/10$, $0<\mu\leq 6$, and $0< \left|z\right|\leq 10$.
\label{Fig8}}
\end{figure}

The Mandel parameter $Q_k\left(\left|z\right|^2\right)$ is displayed in Figures \ref{Fig5} and \ref{Fig6}, for truncated Mittag-Leffler GCSs, and in Figures \ref{Fig7} and \ref{Fig8}, for truncated Wright GCSs, and different values of the dimension, $k+1$, of the truncated Fock space, of the label $z$ and the involved parameters. For small nonvanishing values of the label, the Mandel parameter of truncated Mittag-Leffler GCSs is positive or negative according to the values of the involved parameters, while the Mandel parameter of truncated Wright GCSs is uniquely negative. Instead, for large values of the label, the Mandel parameter is uniquely negative for both the truncated Mittag-Leffler and truncated Wright GCSs. These behaviors are in accordance with the theoretical results.

\subsection{Bunching or anti-bunching effect}\label{42}
The bunching or anti-bunching effects are studied by evaluating the second-order correlation function for the 
 GCSs and truncated GCSs \cite{Truncosc2020},
\begin{eqnarray}
\mathfrak{C}=\frac{\langle \left(a^{\dagger}\right)^2 a^{2}\rangle }
{\langle a^{\dagger} a \rangle^{2}}. \label{Cdef}
\end{eqnarray}
This function is larger (smaller) than unity, $\mathfrak{C}>_{\left(<\right)}1$, in case bunching (anti-bunching) effect appears. In case $k=1$, the corresponding second-order 
correlation function vanishes, $\mathfrak{C}_1\left(\left|z\right|^2\right)=0$, while the Mandel parameter is negative, $Q_1\left(\left|z\right|^2\right)<0$, for every nonvanishing complex value of the label. In this case, the Mandel parameter $Q_1\left(\left|z\right|^2\right)$ departs from the value of the second order correlation function $\mathfrak{C}_1\left(\left|z\right|^2\right)$. In case $k=2,3,\ldots,+\infty$, the truncated GCSs and GCSs, Eq. (\ref{GCoherent1k}) with $k<+\infty$ and $k=+\infty$, respectively, exhibit the following form of the second-order correlation function:
\begin{eqnarray}
&&\mathfrak{C}_k\left(\left|z\right|^2\right)=
\frac{\left[\sum_{n=0}^{k-2}(n+1)(n+2)\left|z\right|^{2n}/g(n+2)\right]\left[\sum_{n=0}^{k}\left|z\right|^{2n}/g(n)\right]}{\left[\sum_{n=0}^{k-1}(n+1)\left|z\right|^{2n}/g(n+1)\right]^2}, 
\label{CkGCSs}
\end{eqnarray}
for every nonvanishing complex value of the label. The following condition about the Mandel parameter: $Q_k\left(\left|z\right|^2\right)>_{\left(<\right)}0$, i.e., the positive (negative) sign, is equivalent to the following relation involving the second-order correlation function: $\mathfrak{C}_k\left(\left|z\right|^2\right)>_{\left(<\right)}1$, for every $k=2,3,\ldots,+\infty$, and for every nonvanishing value of the label.

\section{Summary and conclusions}\label{5}

Truncated GCSs of a quantum harmonic oscillator are defined by requiring, over the finite-dimensional Fock space, the conditions of normalizability, continuity in the label and resolution of the identity operator with a positive weight function. These properties allow to identify GCSs or truncated GCSs such that the corresponding distributions of the number of excitations depart in various ways from the canonical behavior. In fact, for large values of the number of excitations, these distributions decay according to combinations of stretched exponential functions, power laws and powers of logarithmic forms, for both finite- and infinite-dimensional Fock space.

The classical or nonclassical nature of GCSs is investigated by studying the sign of the Mandel $Q$ parameter. For small nonvanishing values of the label $z$, the Mandel $Q$ parameter of the GCS $|z;g \rangle$ is positive or negative, based on the realization of determined constraints which involve the arithmetic function $g(n)$. Mittag-Leffler and Wright GCSs are studied as special cases. For small nonvanishing values of the label $z$, the Mittag-Leffler GCSs exhibit (nonclassical) sub-Poissonian or super-Poissonian statistics of the number of excitations, based on the values of the involved parameters, while the Wright GCSs exhibit uniquely sub-Poissonian statistics of the number of excitations. The second-order correlation function is larger or smaller than unity along with the sign, positive or negative, respectively, of the Mandel $Q$ parameter for every nonvanishing value of the label, both in the finite- and infinite-dimensional case.

In conclusion, CSs describe single-mode quantized states of light which are fundamental in quantum optics due to their properties \cite{Sch1926,Glauber1963,Klauder1963,Klauder1985,Perelomov1986,Klauder1995,Loudon1983,WallsMilbourn1994,MandelWolf1995,SCSsSanders2012}. Particularly, these states are characterized by Poisson photon-counting statistics. GCSs and, particularly, Mittag-Leffler or Wright GCSs describe deviations from canonical CSs. Same property holds for truncated GCSs, and, particularly, for truncated Mittag-Leffler or Wright GCSs, with respect to truncated canonical CSs. Hence, qualitatively, the present approach might help in quantum optics to interpret photon-counting statistics which depart from canonical Poisson distributions in terms of GCSs of light, or truncated GCSs in the finite-dimensional case.

\appendix

\section{Details}\label{A}

Expressions (\ref{pEasympt}) and (\ref{pWasympt}) 
of the probabilities $p_{\alpha,\beta}\left(n, z\right)$, Eq. (\ref{pML}), and $p_{\lambda,\mu}\left(n, z\right)$, Eq. (\ref{pW}), respectively, are obtained in straightforward way from the asymptotic expansion of the involved Gamma function as 
$n \to +\infty$ \cite{AbrHandBook}. By expressing the identity operator in the Fock basis, $I= \sum_{n=0}^{\infty}|n\rangle\langle n|$, the completeness relation (\ref{completnessWCS}) of the Wright GCSs reads as below:
\begin{eqnarray}
\int_{\mathbb{R}^2}U^{\left(W\right)}_{\lambda,\mu}
\left(\left|z\right|^2\right)\langle n|z;\lambda,\mu \rangle\langle z;\lambda,\mu|n^{\prime}\rangle
 d^2 z=\delta_{n,n^{\prime}},
\label{completnessMLCSnnprime}
\end{eqnarray}
for every $n,n^{\prime} \in \mathbb{N}$, where $d^2 z=d\operatorname{Re}z  d\operatorname{Im}z$. This condition becomes 
\begin{eqnarray}
\pi \int_{0}^{\infty}\left[W_{\lambda,\mu}\left(u\right)\right]^{-1}
U^{\left(W\right)}_{\lambda,\mu}
\left(u\right)u^n du=n!\Gamma(\lambda n +\mu),
\label{completnessMLCSnn}
\end{eqnarray}
for every $n\in \mathbb{N}$ and $n^{\prime}=n$. The real variable $u$ is defined as $u=\left|z\right|^2$. Relation (\ref{completnessMLCSnn}) is fulfilled for every natural value $n$ and $\lambda,\mu>0$, if (sufficient condition) the following Mellin transform:  
\begin{eqnarray}
\mathfrak{M}\left[
\pi \left[W_{\lambda,\mu}\left(u\right)\right]^{-1}
U^{\left(W\right)}_{\lambda,\mu}
\left(u\right)\right]\left(s\right)=
\Gamma(s)\Gamma(\lambda s+\mu-\lambda), \hspace{2.4em}
\label{completnessMLCSMellin}
\end{eqnarray}
holds for $\operatorname{Re} s\geq 1$. The Mellin transform $\mathfrak{M}\left[\varphi\left(u\right)\right](s)$, of a general function $\varphi\left(u\right)$, acting over the interval $\left(0,\infty\right)$, is defined by the expression below \cite{TitchmarshFT,Widder,DoetschHLT,MarichevMT},
\begin{eqnarray}
\mathfrak{M}\left[\varphi\left(u\right)\right]\left(s\right)
=\int_0^{\infty} \varphi\left(u\right)u^{s-1}du, \hspace{2.5em}
\label{MellinTdef}
\end{eqnarray}
for every value of the complex variable $s$ such that the above integral exists. Here, the Mellin transform $\mathfrak{M}\left[\varphi\right](s)$ is also labeled $\hat{\varphi}(s)$, for the sake of shortness. The Mellin convolution product \cite{TitchmarshFT,Widder,DoetschHLT,MarichevMT} provides in straightforward way the inverse Mellin transform of the right hand side of Eq. (\ref{completnessMLCSMellin}),
\begin{eqnarray}
&&\mathfrak{M}^{-1}\left[\Gamma(s)
\Gamma(\lambda s+\mu-\lambda)\right](u)=
\lambda^{-1}
\int_0^{\infty}v^{\left(\mu/\lambda\right)-2}
\exp\left(-\frac{u}{v}-v^{1/\lambda}\right)dv \nonumber \\
&&
=\lambda^{-1}Z^{\left(\mu/\lambda\right)-1}_{1/\lambda}(u)=
H^{2,0}_{0,2}\left[u\Bigg|\begin{array}{rr}\\ 
\left(0,1\right)&\left(\mu-\lambda,\lambda\right)
\end{array}\right],\hspace{1.5em}
\label{mellininv1}
\end{eqnarray}
for $\operatorname{Re} \,s> \max \left\{0,1-\left(\mu/\lambda\right)\right\}$, and, particularly, for $\operatorname{Re} \,s \geq 1$, as $1>\max \left\{0,1-\left(\mu/\lambda\right)\right\}$. The integral form appearing in Eq. (\ref{mellininv1}) is a positive function which represents the Kr\"atzel function \cite{Kf}, $Z^{\left(\mu/\lambda\right)-1}_{1/\lambda}(u)$. This function is a special case of the Fox $H$-function 
\cite{FoxHbook}. Hence, expression (\ref{WeightWCS}), obtained in straightforward way from Eqs. 
(\ref{completnessMLCSnn}) and (\ref{mellininv1}), is positive and represents the weight function of the Wright GCSs. 

Expression (\ref{pgnfasympt}) of the 
probability $p_{g}\left(n, z\right)$ is obtained from Eq. (\ref{pgnf}) by performing the asymptotic expansion of the 
function $\hat{f}(s)$ for divergent real values of the variable, $s\to+\infty$ \cite{MellinAsymptSidi1985}. The normalization factor $\mathfrak{N}_1\left(|z|^2\right)$ is obtained from Eq. (\ref{gn1}) by realizing the power series which defines the Mittag-Leffler function. Form (\ref{pgn1asympt}) is found from Eq. (\ref{pgn1}) by considering the 
asymptotic expansion of the Gamma function $\Gamma\left(\alpha n + \beta\right)$ for $n\to+\infty$ \cite{AbrHandBook}. Expressions (\ref{PhiStatecompletnessGCS}) - (\ref{psin}) are obtained from the completeness  property (\ref{completnessGCS}) of the GCSs. 

Expressions (\ref{MLTCkdef})-(\ref{pgnfk}) are obtained in straightforward way by adapting the corresponding canonical forms to the $(k+1)$ finite-dimensional Fock space. Expressions (\ref{pgnfkz0})-(\ref{pgkfkz0}) are obtained from Eq. (\ref{pgnfk}) by performing the asymptotic expansion of the term $\left[\mathfrak{N}_{k,g}\left(|z|^2\right)\right]^{-1}$ 
for $|z|\to 0^+$, with $z\neq 0$. Similarly, forms (\ref{pgnfkzinfty})-(\ref{pgkfkzinfty}) are found by evaluating the asymptotic expansions of the term 
$\left[\mathfrak{N}_{k,g}\left(|z|^2\right)\right]^{-1}$ 
for $|z|\to +\infty$. Forms (\ref{GCoherent11})-(\ref{QTGCSs}) are obtained in straightforward way 
from Eqs. (\ref{GCoherent1k}) and (\ref{Qdef}). Conditions 
(\ref{condg0g1g2QMm0})-(\ref{condg0g3g1g2Qeq0})
are obtaind from the asyptotic expansion of expression 
(\ref{QTGCSs}) of the mandel $Q$ parameter for $\left|z\right|\to 0^+$, with $z\neq 0$. Form (\ref{Qkinfty}) is found from the asymptotic expansion of expression (\ref{QTGCSs}) for 
$\left|z\right|\to +\infty$. Expression (\ref{CkGCSs}) is obtained in straightforward way from Eqs. (\ref{Cdef}) and (\ref{GCoherent1k}). This concludes the demonstration of the present results.

\acknowledgments{The research activity of the authors has been carried out in the framework of the activities of the National Group of Mathematical Physics (GNFM, INdAM).}

\end{document}